\documentstyle[11pt,aaspp4]{article}

\lefthead{Verner et al.}
\righthead{Fits for Photoionization Cross Sections} 

\begin{document}
 
\title{Atomic Data for Astrophysics. II.
New Analytic Fits for Photoionization Cross Sections of Atoms and Ions}

\author{D. A. Verner\altaffilmark{1}, G. J. Ferland, and K. T. Korista}
\affil{Department of Physics and Astronomy, University of Kentucky,
Lexington, KY 40506}

\and

\author{D. G. Yakovlev}
\affil{Ioffe Institute of Physics and Technology, St. Petersburg 194021, 
Russia}

\altaffiltext{1}{verner@pa.uky.edu}

\begin{abstract}
We present a complete set of analytic fits to the non-relativistic
photoionization cross 
sections for the ground states of atoms and ions of elements from H through 
Si, and S, Ar, Ca, and Fe. Near the ionization thresholds, the fits are based 
on the Opacity Project theoretical cross sections interpolated and smoothed 
over resonances. At higher energies, the fits reproduce calculated 
Hartree-Dirac-Slater photoionization cross sections. 
\end{abstract}


\section{Introduction}

Photoionization cross sections of atoms and ions are needed for many 
astrophysical applications. In earlier papers \markcite{vern1993}
(Verner et al. 1993, hereafter
Paper I; \markcite{vern1995}
Verner \& Yakovlev 1995, hereafter Paper II) we presented the 
Hartree-Dirac-Slater (HDS) calculations and analytic fits to the partial 
photoionization cross sections for all subshells $nl$ of the ground states of 
all atoms and ions from H to Zn. Comparison with available experimental and 
theoretical data made in Paper II demonstrated the high accuracy of the cross 
section fits for inner shells and far from the thresholds of outer shells. 
However, the theoretical HDS cross sections can be inaccurate near the 
thresholds of the outer shells, especially for neutral and low ionized species.
The recent Opacity Project (OP) calculations \markcite{seat1992}
(Seaton et al. 1992) based on the 
higher level methods such as the R-matrix method give more accurate low energy
photoionization cross sections. 

In this paper we present a set of analytic fits to the photoionization cross 
sections, interpolated and smoothed over resonances, for the ground states of 
atoms and all ions of the OP elements (H, He, Li, Be, B, C, N, O, F, Ne, Na, 
Mg, Al, Si, S, Ar, Ca, Fe). Our fits are valid in the energy range from the 
ionization threshold up to the first cross section jump due to inner shell 
photoionization. In combination with the earlier fits to the inner shell cross 
sections (Paper II), we obtain an accurate and complete set of 
non-relativistic photoionization cross sections for calculating the ionization 
balance in astrophysical plasmas.  

\section{Data}

We have retrieved the ground-state photoionization cross sections for all
the OP species from a database TOPbase version 0.7 \markcite{cunt1993}
(Cunto et al. 1993). 
The OP calculations reproduce 
many autoionization resonances near the ionization thresholds. For many 
practical 
purposes, it is often sufficient to use the photoionization cross sections
smoothed over the resonances (except for some very wide resonances). 
Autoionization
resonances 
could be important for calculating the photoionization rate produced by
radiation in spectral lines whose positions coincide with the autoionization
resonances. 
However, the OP theoretical energy levels deviate from experimental ones
by 1--2\% \markcite{vern1994} (Verner, Barthel, \& Tytler 1994). 
This accuracy is clearly insufficient for
astrophysical spectroscopy, because the uncertainty in the position is greater
than the width of the resonances.
If one would use the original OP photoionization cross 
sections for calculating the photoionization rates produced by radiation in 
spectral lines, one could obtain occasional coincidences and/or disagreements
between spectral line energies and the OP resonances which do not occur in 
nature. Therefore, smoothing of
the OP photoionization cross sections is preferable.

For ions with 
a number of electrons $5 \le N \le 10$, the OP calculations give the total 
photoionization cross section from $2p+2s$ shells. The HDS calculations show a 
jump in the total cross section at the $2s$ shell ionization threshold.
However, more advanced R-matrix calculations reveal a series of autoionization 
resonances approaching the $2s$ threshold. After smoothing over the resonances,
the jump in the total cross section disappears. Experimental data available 
(see, e.g., \markcite{sams1990} 
Samson \& Angel 1990) do not show such jumps either. However,
a large cross section jump at the subsequent $1s$ ionization threshold is 
observable. This jump at the $1s$ shell edge is real due to a large energy 
difference between the $1s$ and $2s$ shell ionization potentials. Therefore,
it is reasonable to give a fit to the total ($2p+2s$) photoionization cross 
section from the ionization threshold up to the $1s$ shell edge, and to 
supplement it by a separate fit to the $1s$ shell photoionization cross section
at energies above the $1s$ edge energy. Note that the inner $1s$ shell 
ionization leads to subsequent ejection of Auger electrons. 

A similar situation
occurs for the $3p$ and $3s$ photoionization cross sections of the third-row 
elements with $13 \le N \le 18$. The OP data present the total cross sections, 
and there are no jumps at the $3s$ ionization thresholds. For these species, 
we give the fits to the total ($3p+3s$) cross sections. The photoionization 
from inner $2p$, $2s$, and $1s$ shells is followed by ejection of Auger 
electrons. The partial photoionization cross sections from these shells can be 
taken from Paper II; the probabilities of corresponding Auger processes are
listed by \markcite{kaas1993} Kaastra \& Mewe (1993).

The R-matrix OP calculations have been done at energies below the largest 
energy of the target states in the close-coupling expansion. At higher 
energies, the TOPbase gives simple power-law tails. Comparison with experiment
made in Paper II shows that these tails are inaccurate. They have been excluded
from our consideration. For each species, the low-energy OP data were 
interpolated, and the data set with a constant energy step ($\leq$ the smallest
step in the original OP data) was created. This data set was smoothed until all
resonances disappeared. In most of the cases, this happened at an energy step
$\Delta E/E < 2-3\%$; which is comparable with the accuracy of the OP energy 
calibration. In some cases, very big resonances cannot be smoothed at any 
reasonable energy step. In these cases, our fits match the effective background
determined from smoothing of all smaller resonances. 

For most of the species, the smoothed OP data at the largest energies 
calculated by the OP are in a very good agreement with the fit to the total 
photoionization cross section from Paper II. Discrepancies are usually less 
than 2\%. In these cases, we simply add the HDS data to the smoothed OP data at
higher energies to trace the cross section up to the subsequent inner-shell 
ionization edge. The situation is more complicated for neutrals and singly 
ionized species with outer $3p$ electron. The HDS data do not match the OP 
data at the highest OP energy below the power-law tail.
Experimental data for these species are
available in the case of \ion{Ar}{1}. Figure~1 shows the recommended
experimental photoionization cross section of \ion{Ar}{1} \markcite{marr1976}
(Marr \& West 1976).
Both the experimental and the OP data show the cross section maximum near the 
threshold. Both the experimental and the HDS data show the cross section
minimum at a higher energy, and then the second cross section maximum. The OP 
data are available only at energies below the position of the cross section 
minimum. The absolute value of the cross section minimum is in reasonable 
agreement with the HDS calculation, but the HDS data do not give an energy 
position of the minimum. The HDS data are in a good agreement with the 
experiment at energies above the energy of the second cross section maximum.
For fitting such cases, we have used the smoothed OP data supplemented by the 
HDS data above the second maximum. We have also taken into account the absolute
value of the cross section minimum. For \ion{Ar}{1}, we fit the recommended 
experimental data. We also give a fit to the smoothed experimental data for the
$4s$ shell photoionization cross section of \ion{Ca}{2} \markcite{lyon1987}
(Lyon et al. 1987),
since the TOPbase gives only a simple power-law tail for this ion. The TOPbase 
does not include the data for \ion{Fe}{1} and \ion{Fe}{2}. For these species, 
we have used the smoothed total ($4s+3d$) photoionization cross sections taken 
from the R-matrix calculations by \markcite{naha1994} 
Nahar \& Pradhan (1994) and \markcite{baut1995} Bautista \& 
Pradhan (1995). For \ion{He}{1}, we give a fit to highly accurate experimental
absolute photoionization cross section measured by \markcite{sams1994}
Samson et al. (1994).

\section{Fits to the photoionization cross sections}

We propose to describe the photoionization cross sections $\sigma(E)$ from the 
outer shells of atoms and ions in question (Sect.~2) by the fitting formula:
\begin{eqnarray}
    \sigma(E)=\sigma_0 \, F(y) \;\; {\rm Mb}, 
 \;\;\;\;\;\; 
     x = {E \over E_0} -y_0, 
 \;\;\;\;\;\; 
     y = \sqrt{x^2+ y_1^2}, 
\nonumber\\
     F(y) = \left[(x-1)^2+y_{\rm w}^2 \right]y^{0.5P-5.5}
          \left(1+ \sqrt{y/y_{\rm a}} \right )^{-P},
    \label{eq:1}
\end{eqnarray}
where $E$ is photon energy in eV, and $\sigma_0$, $E_0$, $y_{\rm w}$,
$y_{\rm a}$, $P$, $y_0$ and $y_1$ are the fit parameters
(1~Mb = $10^{-18}$~cm$^2$). Equation (\ref{eq:1})
generalizes the fitting formula used in Papers I and II, and it reduces to the 
formula of Papers I and II if $y_0=y_1=0$. Additional fitting parameters $y_0$ 
and $y_1$ are required to describe adequately the more complicated
behavior of the smoothed OP data near the thresholds 
of the outer shells of neutral and low ionized species, including the 
two-maxima cross section structure for the third row elements. 

The fit parameters $E_0$ and $\sigma_0$ determine, respectively, typical energy
and cross section scales (see Papers I and II for details). Other fit 
parameters reflect the cross section features. In particular $y$ can be 
rewritten as $y= \sqrt{[(E-E_1)^2 + \Gamma_1^2]/E_0}$, where $E_1=y_0 \, E_0$ 
shows the position of the first (after the threshold) cross section maximum, 
and $\Gamma_1 = y_1 \, E_0$ is an `energy width' of this maximum. Similarly,
$(x-1)^2+y_{\rm w}^2 = [(E-E_2)^2 + \Gamma_2^2]/E_0^2$, where $E_2=E_1+E_0$ 
indicates the energy position of the cross section minimum (dip), and 
$\Gamma_2= y_{\rm w} \, E_0$ can be treated as the `dip energy width'. The fit 
parameter $P$ determines the power-law cross section behavior
$\sigma(E) \propto E^{0.5P-3.5}$ in the energy interval
$E_2 \ll E \ll E_{\rm a}$ behind the dip (provided $E_{\rm a} \gg E_2$),
where $E_{\rm a} = y_{\rm a} \, E_0$ is the upper boundary of this interval.
Finally Eq.~(\ref{eq:1}) reproduces the correct non-relativistic high-energy 
asymptote, $\sigma(E) \propto E^{-3.5}$, for $E \gg E_{\rm a}$. Note that all 
of the cross sections to be fitted (Sect.~2) include the contribution of the 
outer $s$ subshell at high $E$. If we considered the partial photoionization 
cross sections for another subshell $l$, Eq.~(\ref{eq:1}) should have been 
modified by replacing $0.5P-5.5$ with $0.5P-5.5-l$ to ensure the correct 
high-energy asymptote (see Papers I and II).

Equation~(\ref{eq:1}) can be generalized further to describe more complicated 
near-threshold behavior of the photoionization cross sections smoothed over 
resonances. For instance, the smoothed ($3p+3s$)-shell ionization cross section
of \ion{S}{1} looks as if another minimum occurs very near
the outer 3$p$-shell ionization threshold. This conclusion is 
not definite since there are many autoionization near-threshold resonances.
If, however, this were so, we could include the effect into the fitting by
introducing into (\ref{eq:1}) an additional factor
$[(E-R_3)^2+\Gamma_3^2]/E^2$, where $E_3$ and $\Gamma_3$ are two new fit 
parameters which determine, respectively, the position and energy width of the 
new dip. We do not introduce this improvement into the present results since
our `standard' formula (\ref{eq:1}) gives satisfactory accuracy even for 
\ion{S}{1}. This generalization could be helpful in the future for fitting new 
cross sections.

All the fitting parameters and experimental ionization threshold energies
$E_{\rm th}$ taken from \markcite{kell1987} 
Kelly (1987) are listed in Table~1. 
The fits are obtained
for energies below the energy of the subsequent inner shell ionization edge. 
For species with $3 \le N \le 10$, this is the $1s$ shell ionization energy; 
for $11 \le N \le 18$ -- the $2p$ shell ionization energy; for  
$19 \le N \le 26$ -- the $3p$ shell ionization energy. These energies, taken 
from Paper II, are listed as $E_{\rm max}$ in Table 1. At $E \ge E_{\rm max}$,
the present fits should be supplemented by the fits to the inner shell 
photoionization cross sections given in Paper II. For H-like and He-like 
species ($N \le 2$), the only ground state shell is the $1s$ shell. Since the 
present fitting formula has a correct non-relativistic asymptote, our fits are 
accurate at all non-relativistic energies. Thus, we put $E_{\rm max} = 50$~keV 
for the H-like and He-like species. Note that for the H-like, He-like (except 
\ion{He}{1}) and Na-like (except \ion{Na}{1}) species, which do not show 
resonances in the OP calculations, we keep the fits to the cross sections from 
Paper II. 

All the fits can be divided to several classes of accuracy listed here
starting with the most accurate:
\begin{enumerate}
\item[A.] Hydrogenic cross sections. They are known
exactly, and rms accuracy of the fits is better than 0.2\%.
\item[B.] Cross sections without near-threshold resonances (He-like, Li-like,
and Na-like species).
\item[C.] Regular cross sections. In these data,
smoothed resonances reliably match the background cross section.
\item[D.] The same as C but for neutral and low-ionized species with 
$N \ge 13$. They are less accurate
than cross sections of class C (see discussion in Section 2).
\item[E.] Cross sections with broad near-threshold resonances.
\item[F.] The OP cross sections calculated with insufficient number of
target states in the close-coupling expansion.
\end{enumerate}
Some comments concerning the fits to experimental
cross sections
deserve mention. The \ion{He}{1} data by \markcite{sams1994}
Samson et al. (1994) 
have the 
estimated accuracy
better than 2\% in low-energy region, and better than 10\% in high-energy
region, and our fit does not lead to any loss of accuracy, keeping in
mind that it matches the smoothed cross section in the region of resonances.
The \ion{Ar}{1} experimental data recommended by \markcite{marr1976}
Marr \& West (1976) have
lower accuracy (see \markcite{sams1991}
Samson et al. 1991), and contain uncertainties as
high as 30\%. The \ion{Ca}{2} experimental data by \markcite{lyon1987}
Lyon et al. (1987) include
several huge resonances which cannot be smoothed at any reasonable energy
step.

As an illustration, Figures 2--7 compare the present fits with the OP and 
\markcite{reil1979}
Reilman \& Manson (1979) data and also with the fits to the HDS cross sections 
from Paper II for the neutral, singly ionized and double ionized atoms of the 
astrophysically important elements C ($Z=6$) and Si ($Z=14$). At high energies,
the present fits coincide with those from Paper II.

Concluding, the present fits are generally accurate at low energies, where 
they are
based on the smoothed
OP or experimental
photoionization cross sections, and at high energies where the fits are
based on the HDS
photoionization cross sections. 
The fit parameters in electronic form and a Fortran subroutine which implements
these fits and fits from the Paper II are available through anonymous ftp at 
asta.pa.uky.edu, 
cd~dima/photo, or through the World Wide Web page ``Atomic Data for 
Astrophysics'', http://www.pa.uky.edu/$\sim$verner/atom.html.
The fits will be updated as soon as new, more accurate photoionization
cross sections become available.

\acknowledgements{This work was supported by grants from the National
Science Foundation (AST 93-19034) and NASA (NAGW 3315).
We are grateful to M. J. Seaton for useful discussions on the Opacity Project 
data. We are also grateful to the Opacity Project team for their efforts
to make the data available before publication through the TOPbase database.
We acknowledge the use of the TOPbase version 0.7 database
installed by A. K. Pradhan at the Ohio State University. We thank the referee,
Rolf Mewe, for suggestions which improved the presentation of the paper.
We also thank D. Cohen for his helpful comments.

\appendix
\section{References to the Opacity Project photoionization cross sections}
\begin{tabular}{ll}
He-like & Fernley, Taylor, \& Seaton 1987\\
Li-like & Peach, Saraph, \& Seaton 1988\\
Be-like & Tully, Seaton, \& Berrington 1990\\
\ion{B}{1} & Berrington \& Hibbert, to be published\\
B-like ions & Fernley, Hibbert, Kingston, \& Seaton, to be published\\
C-like & Luo \& Pradhan 1989\\
N-like & Burke \& Lennon, to be published\\
O-like, F-like & Butler \& Zeippen, to be published\\
Ne-like & Hibbert \& Scott 1994\\
Na-like & Taylor, to be published\\
Mg-like & Butler, Mendoza, \& Zeippen 1993\\
Al-like & Mendoza, Eissner, Le Dourneuf, \& Zeippen 1995\\
Si-like & Nahar \& Pradhan 1993\\
P-like, S-like, Cl-like & Butler, Mendoza \& Zeippen, 
to be published\\
\ion{Ar}{1}, \ion{Ca}{1}--\ion{Ca}{3} & Berrington, Hibbert \& Scott, 
to be published\\
\ion{Fe}{3}--\ion{Fe}{4} & Sawey \& Berrington 1992\\
\ion{Fe}{5} & Butler, to be published\\
\ion{Fe}{6}, \ion{Fe}{9}--\ion{Fe}{13} & Mendoza, to be published\\
\ion{Fe}{7}, \ion{Fe}{8} & Saraph, Storey, \& Taylor 1992\\
\end{tabular}

\clearpage

\begin{figure}
\plotfiddle{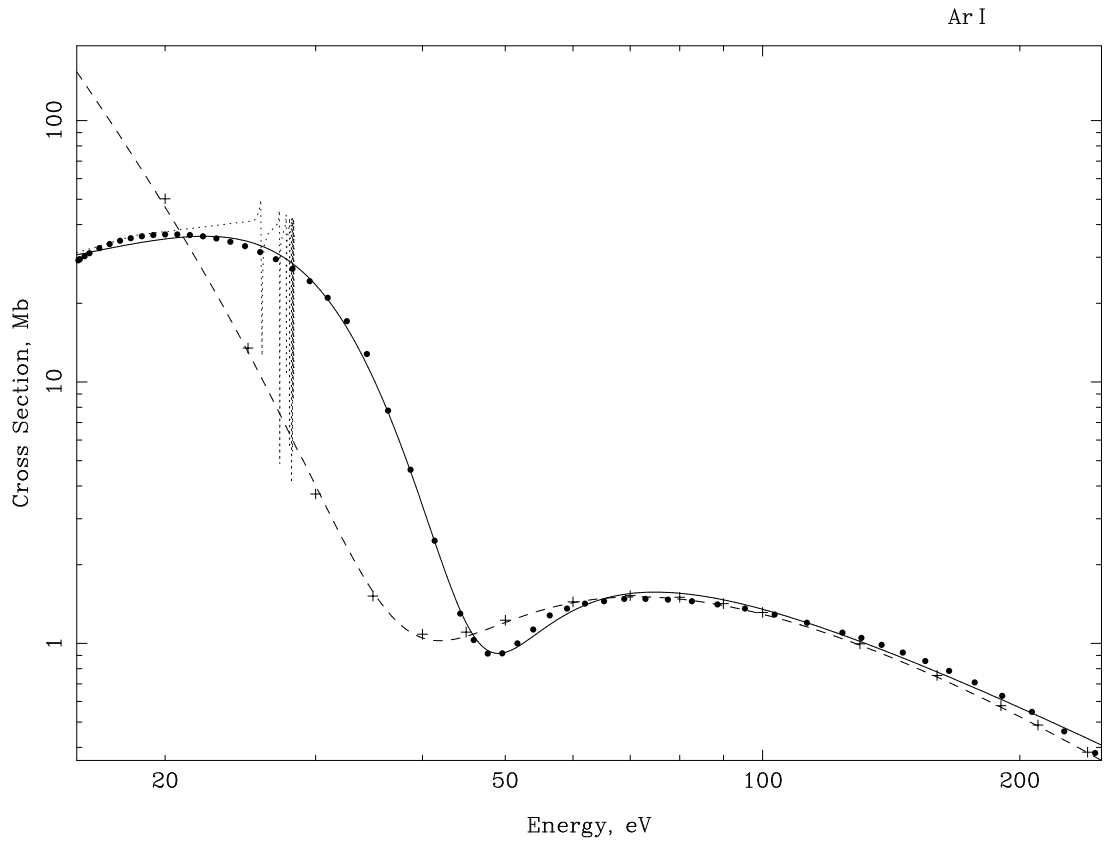}{10cm}{-90}{60}{60}{-230}{350}
\caption{Total photoionization cross
section of Ar~I. Solid line: present work; dashed line: fit from
Paper II; dotted line: Opacity Project data; circles:
recommended experimental data from
Marr \& West (1976); crosses: calculations from Reilman \& Manson (1979).}
\end{figure}
 
\setcounter{figure}{1}
\begin{figure}
\plotfiddle{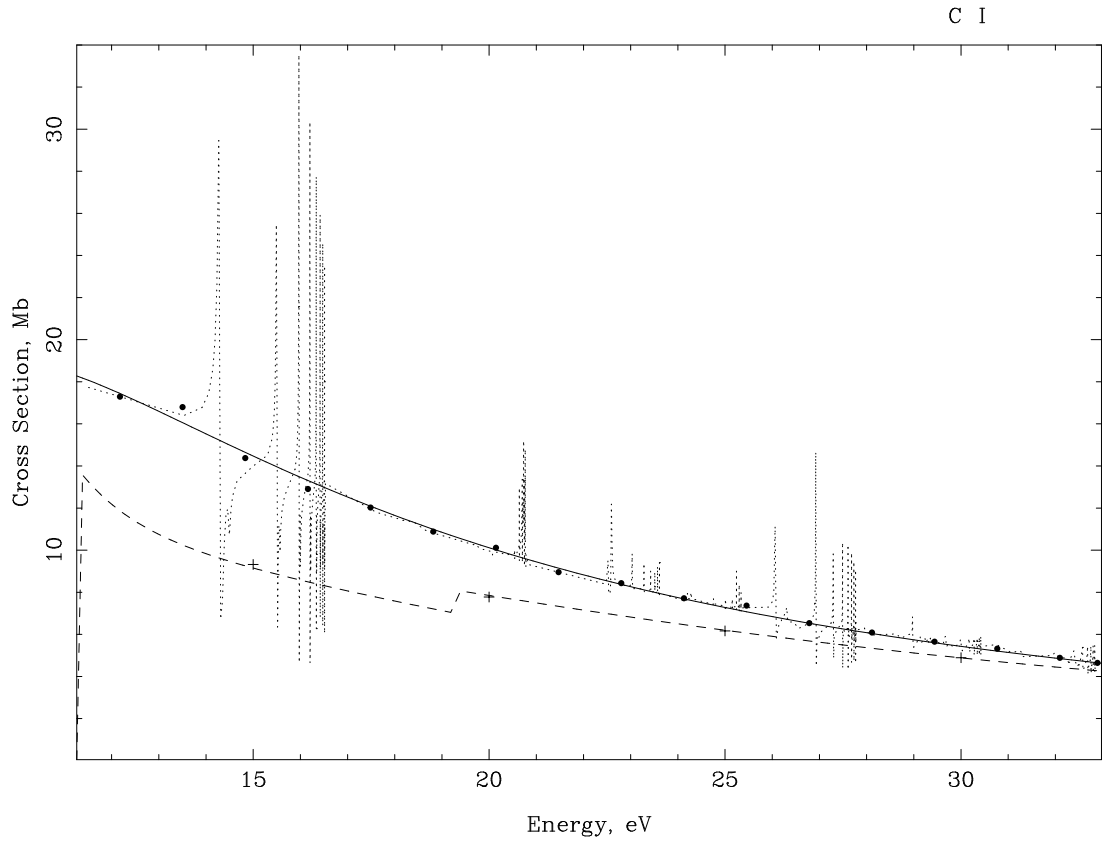}{10cm}{-90}{60}{60}{-230}{350}
\caption{$a$. Total photoionization cross sections of C~I. 
Solid line: present work;
dashed line: fit from Paper II; dotted line: Opacity Project data; circles: 
Opacity Project data interpolated and smoothed over resonances; crosses: 
calculations from Reilman \& Manson (1979). The cross sections are shown in 
linear scale for the Opacity Project energy range.}
\end{figure}

\setcounter{figure}{1}
\begin{figure}
\plotfiddle{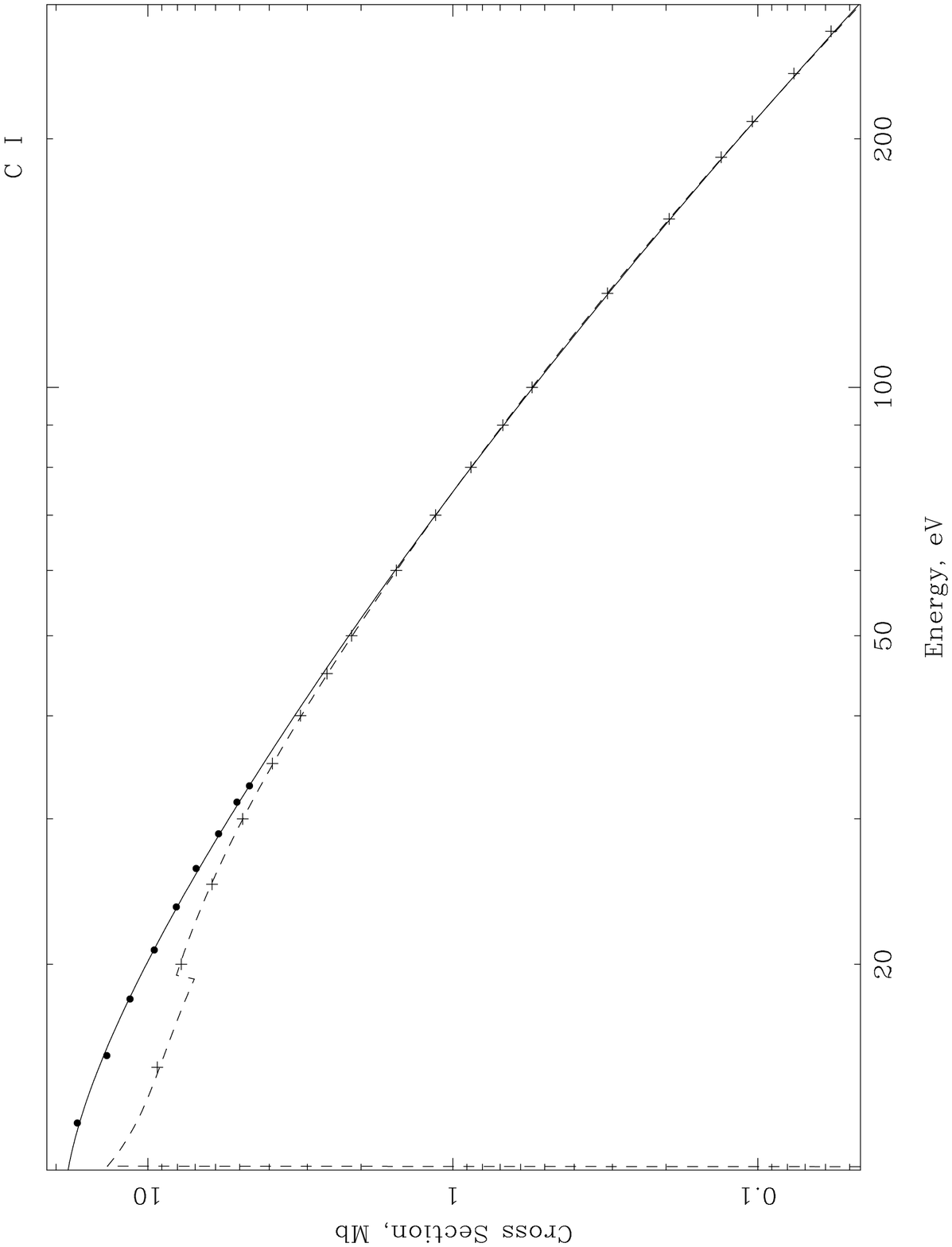}{10cm}{-90}{60}{60}{-230}{350}
\caption{$b$. Same as Fig. 2$a$. 
The cross sections are shown in 
logarithmic
scale from the threshold up to the $1s$ ionization edge.}
\end{figure}

\setcounter{figure}{2}
\begin{figure}
\plotfiddle{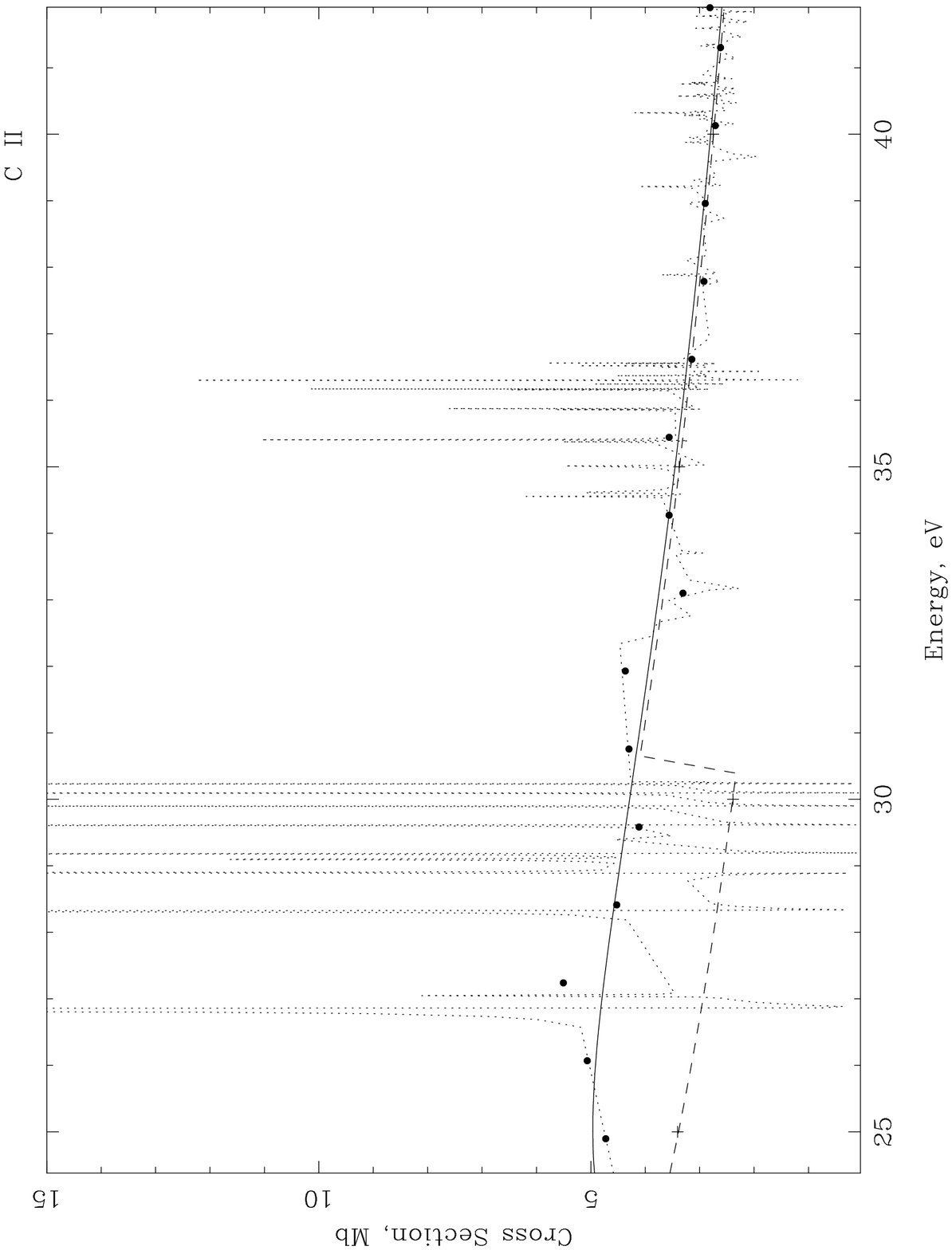}{10cm}{-90}{60}{60}{-230}{350}
\caption{$a$. Same as Fig. 2$a$ for C~II.}
\end{figure}

\setcounter{figure}{2}
\begin{figure}
\plotfiddle{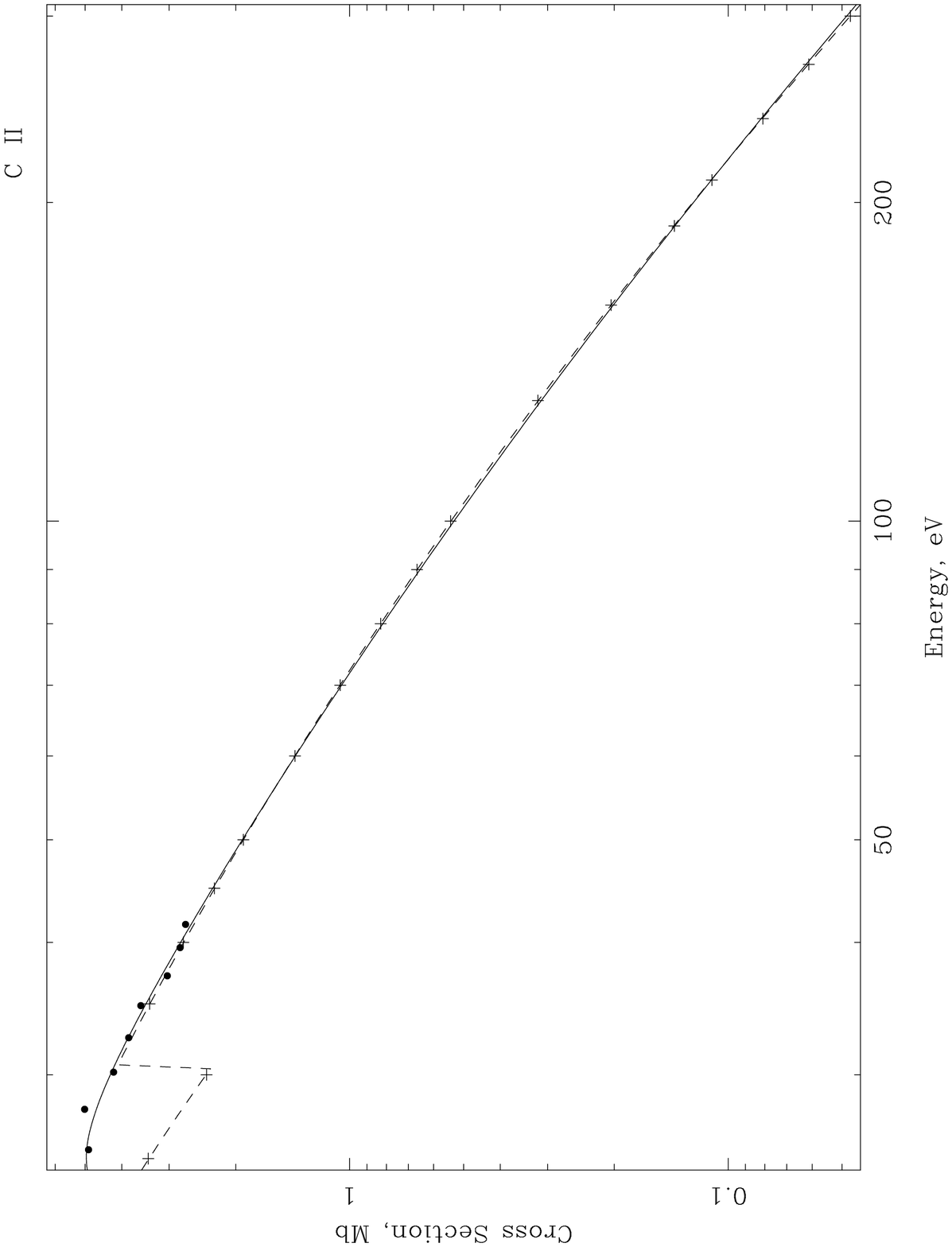}{10cm}{-90}{60}{60}{-230}{350}
\caption{$b$. Same as Fig. 2$b$ for C~II.}
\end{figure}

\setcounter{figure}{3}
\begin{figure}
\plotfiddle{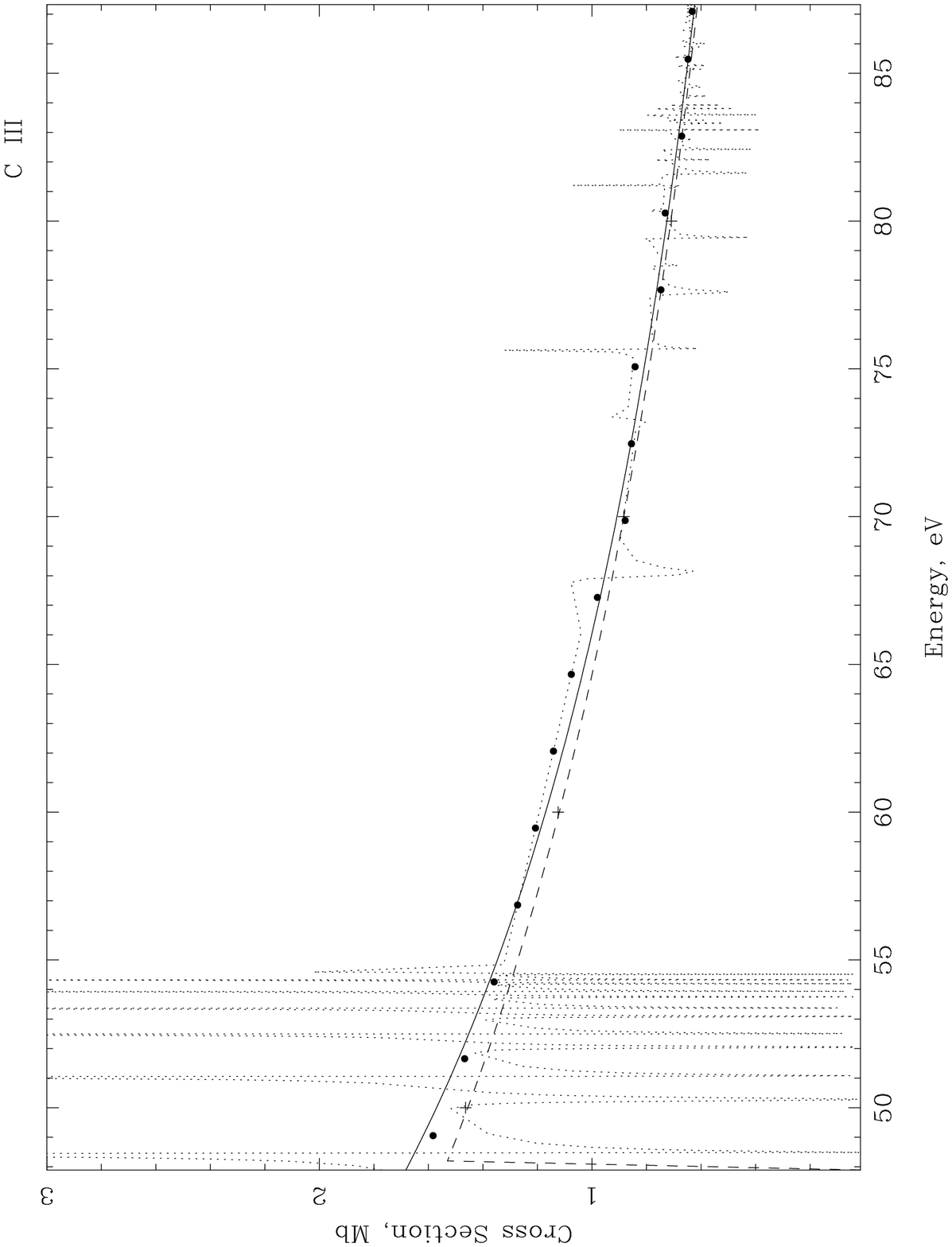}{10cm}{-90}{60}{60}{-230}{350}
\caption{$a$. Same as Fig. 2$a$ for C~III.}
\end{figure}

\setcounter{figure}{3}
\begin{figure}
\plotfiddle{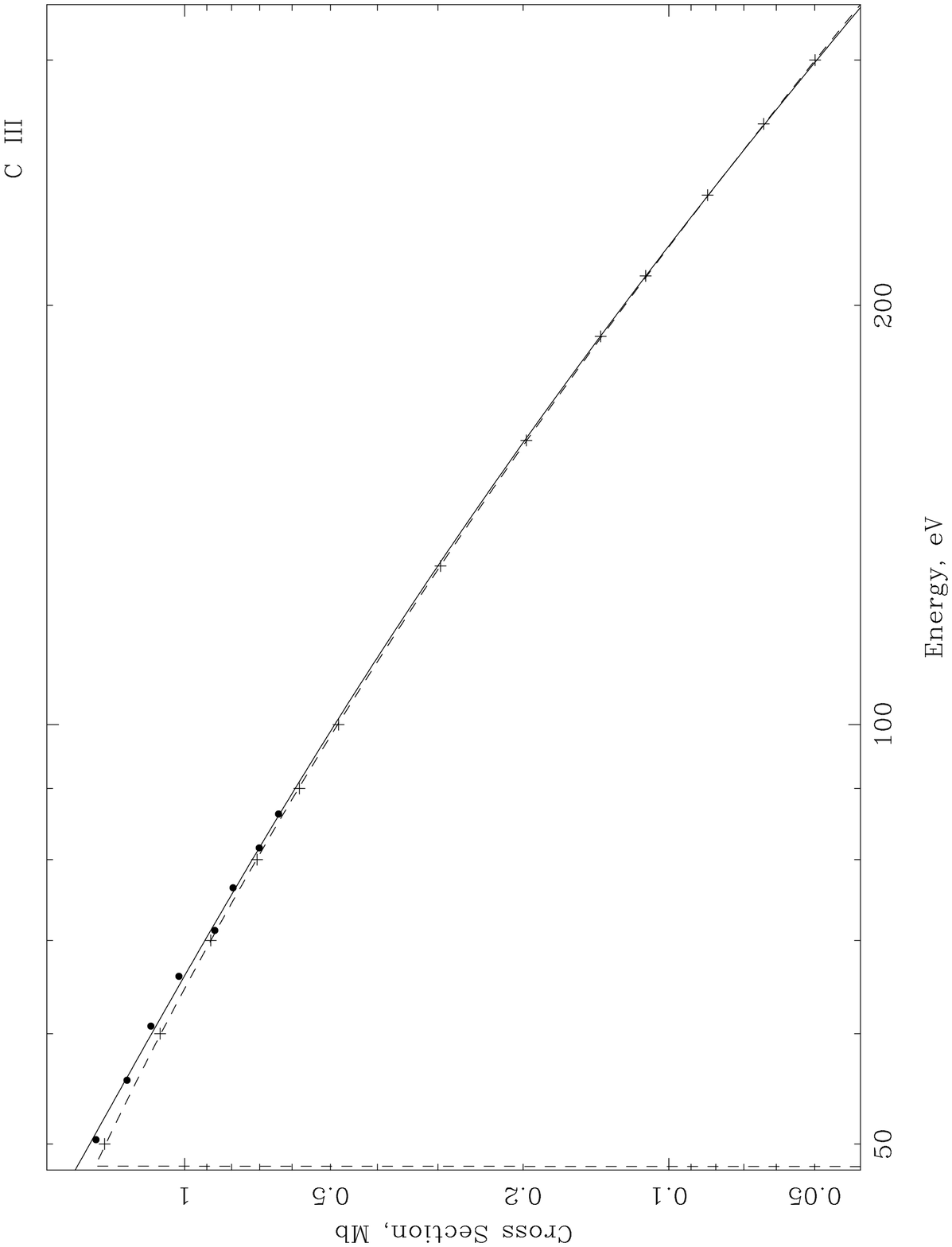}{10cm}{-90}{60}{60}{-230}{350}
\caption{$b$. Same as Fig. 2$b$ for C~III.}
\end{figure}

\setcounter{figure}{4}
\begin{figure}
\plotfiddle{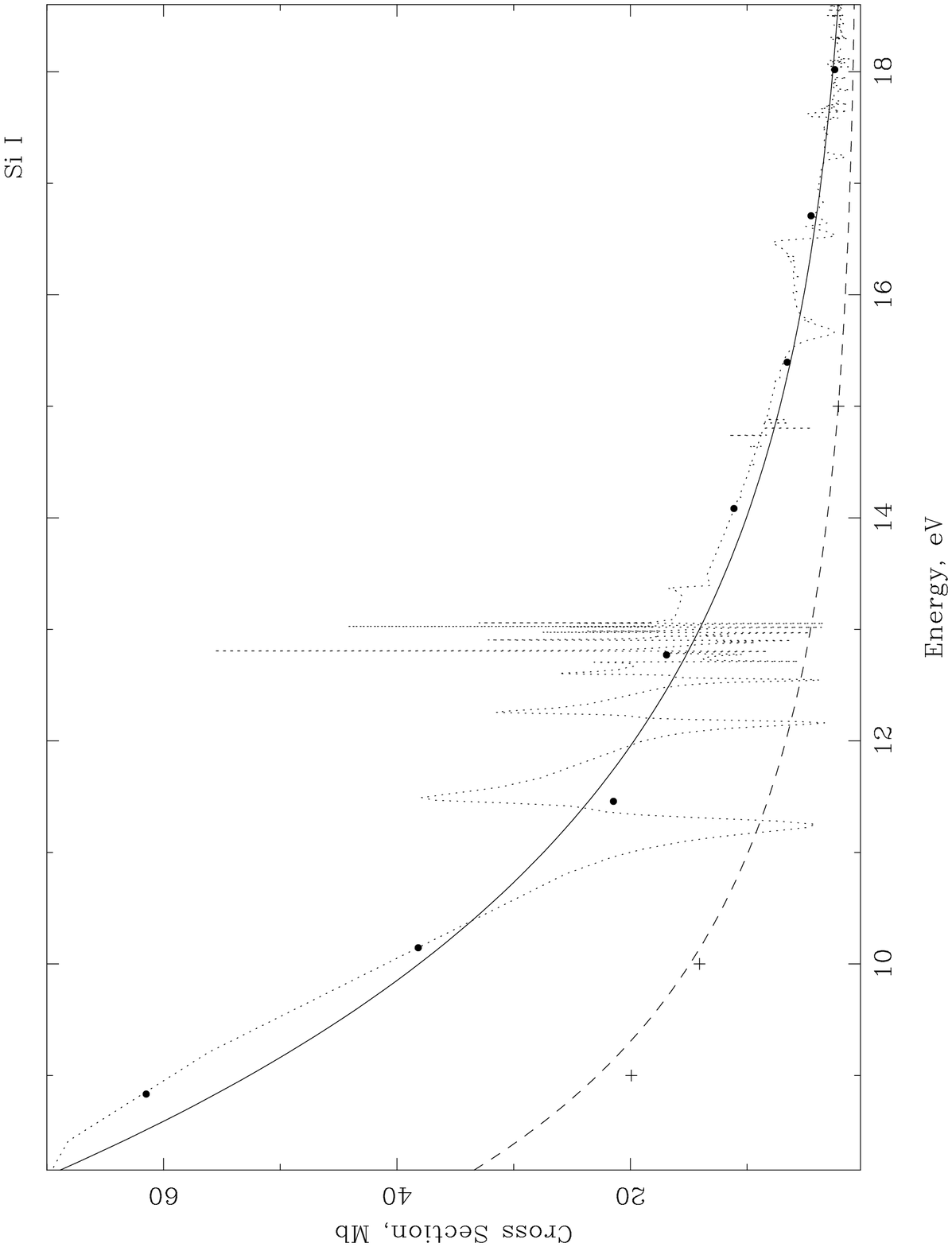}{10cm}{-90}{60}{60}{-230}{350}
\caption{$a$. Same as Fig. 2$a$ for Si~I.}
\end{figure}

\setcounter{figure}{4}
\begin{figure}
\plotfiddle{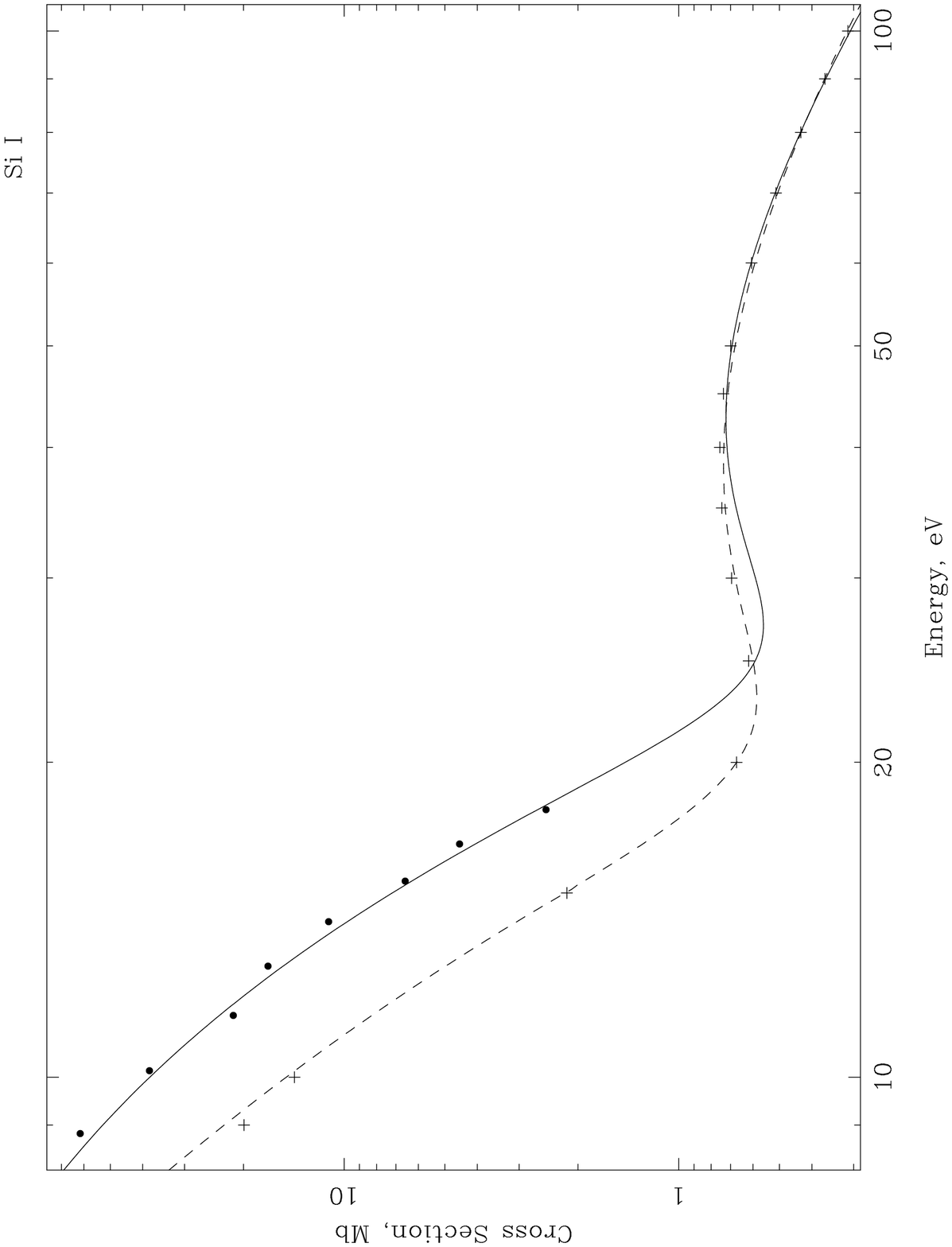}{10cm}{-90}{60}{60}{-230}{350}
\caption{$b$. Same as Fig. 2$b$ for Si~I. 
The cross sections are shown in 
logarithmic scale from 
the threshold up to the $2p$ ionization edge.}
\end{figure}

\setcounter{figure}{5}
\begin{figure}
\plotfiddle{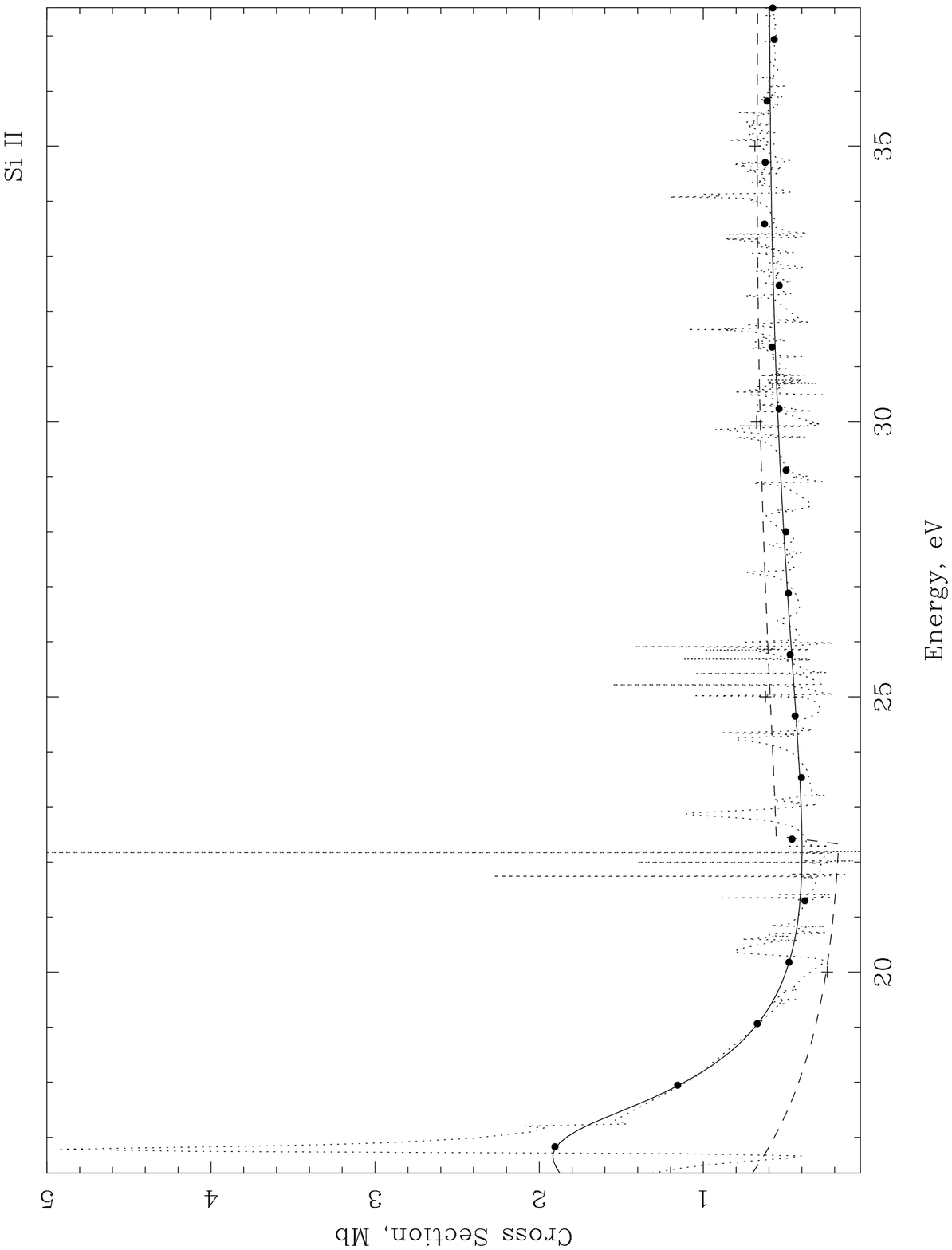}{10cm}{-90}{60}{60}{-230}{350}
\caption{$a$. Same as Fig. 2$a$ for Si~II.}
\end{figure}

\setcounter{figure}{5}
\begin{figure}
\plotfiddle{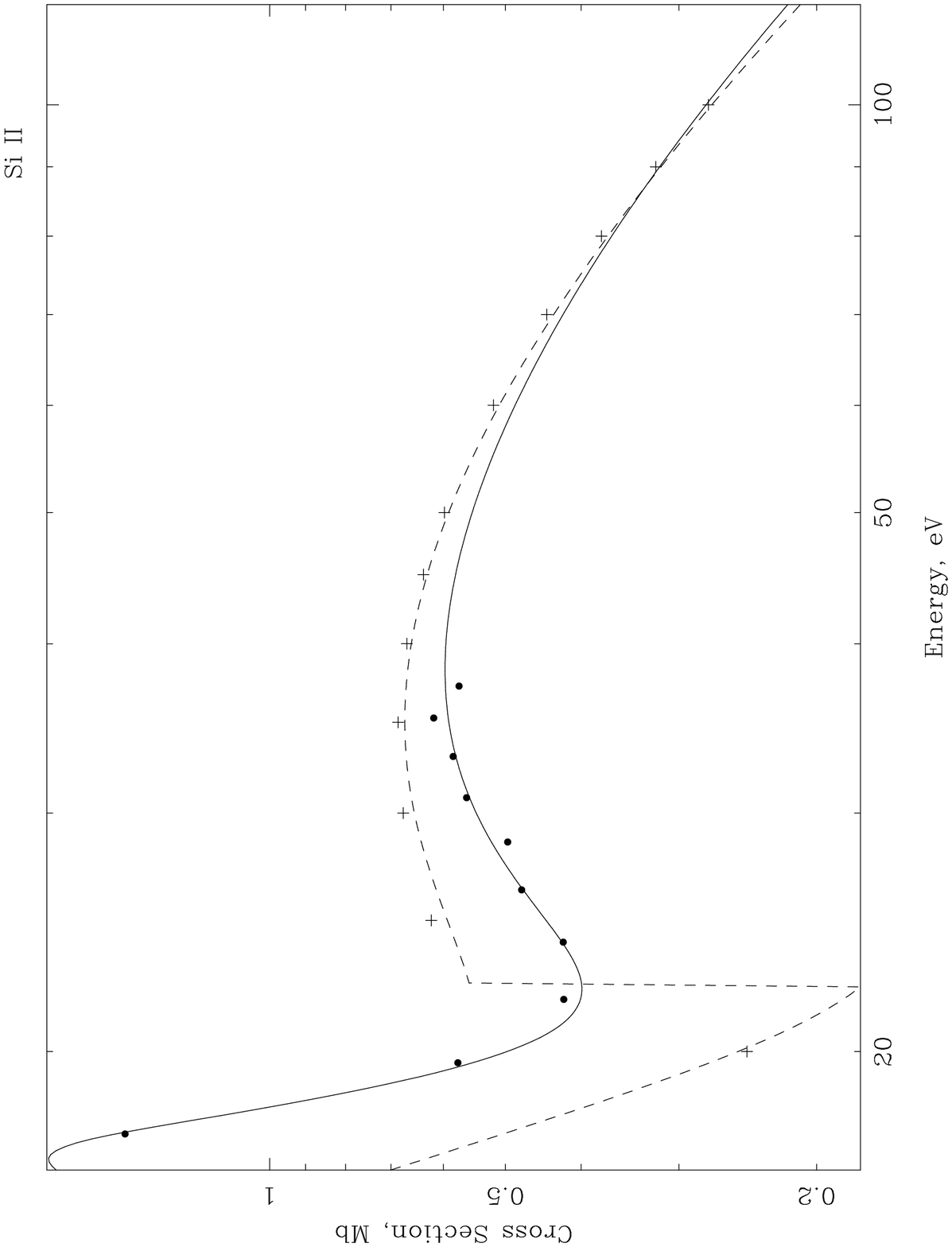}{10cm}{-90}{60}{60}{-230}{350}
\caption{$b$. Same as Fig. 5$b$ for Si~II.}
\end{figure}

\setcounter{figure}{6}
\begin{figure}
\plotfiddle{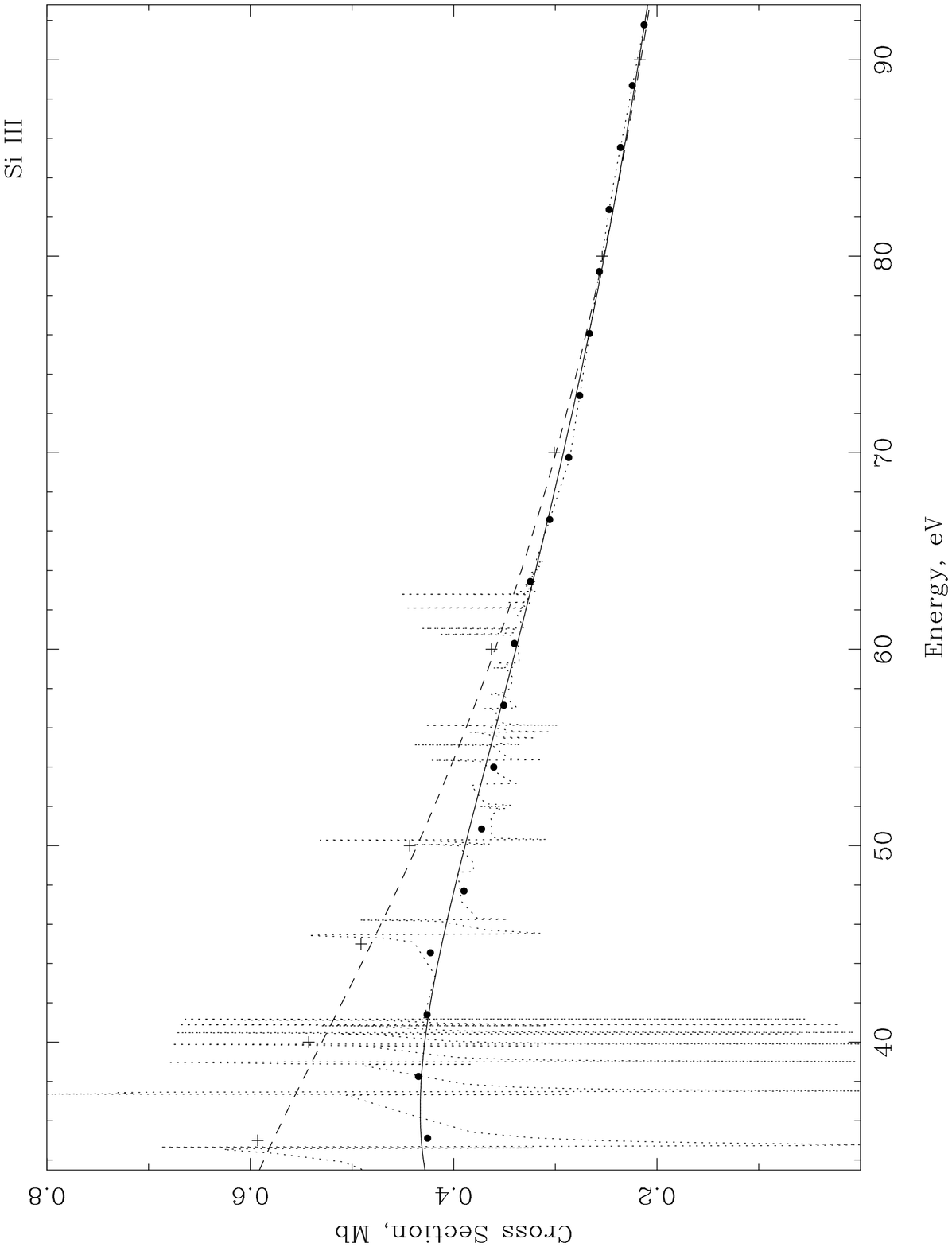}{10cm}{-90}{60}{60}{-230}{350}
\caption{$a$. Same as Fig. 2$a$ for Si~III.}
\end{figure}

\setcounter{figure}{6}
\begin{figure}
\plotfiddle{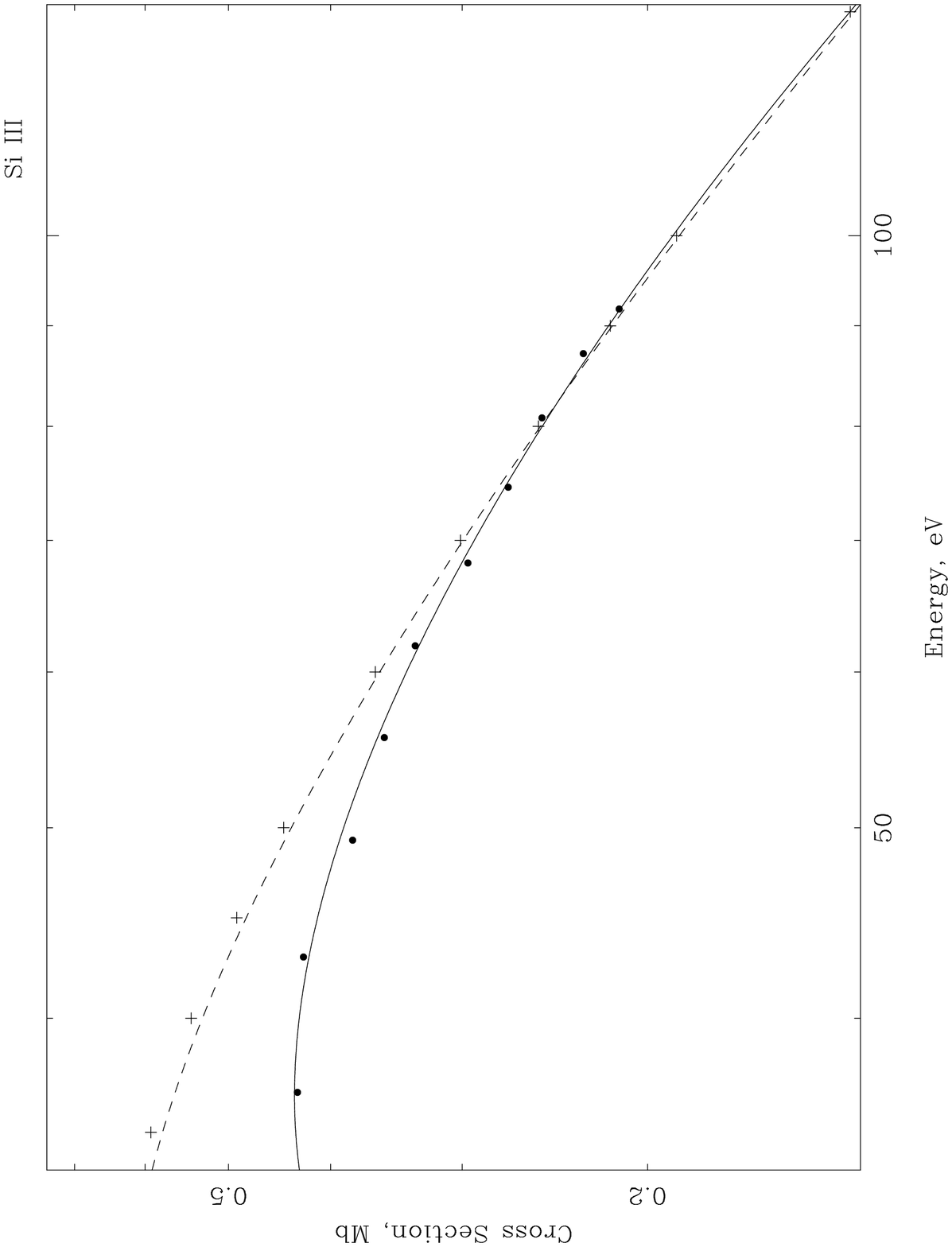}{10cm}{-90}{60}{60}{-230}{350}
\caption{$b$. Same as Fig. 5$b$ for Si~III.}
\end{figure}

\begin{figure}
\plotone{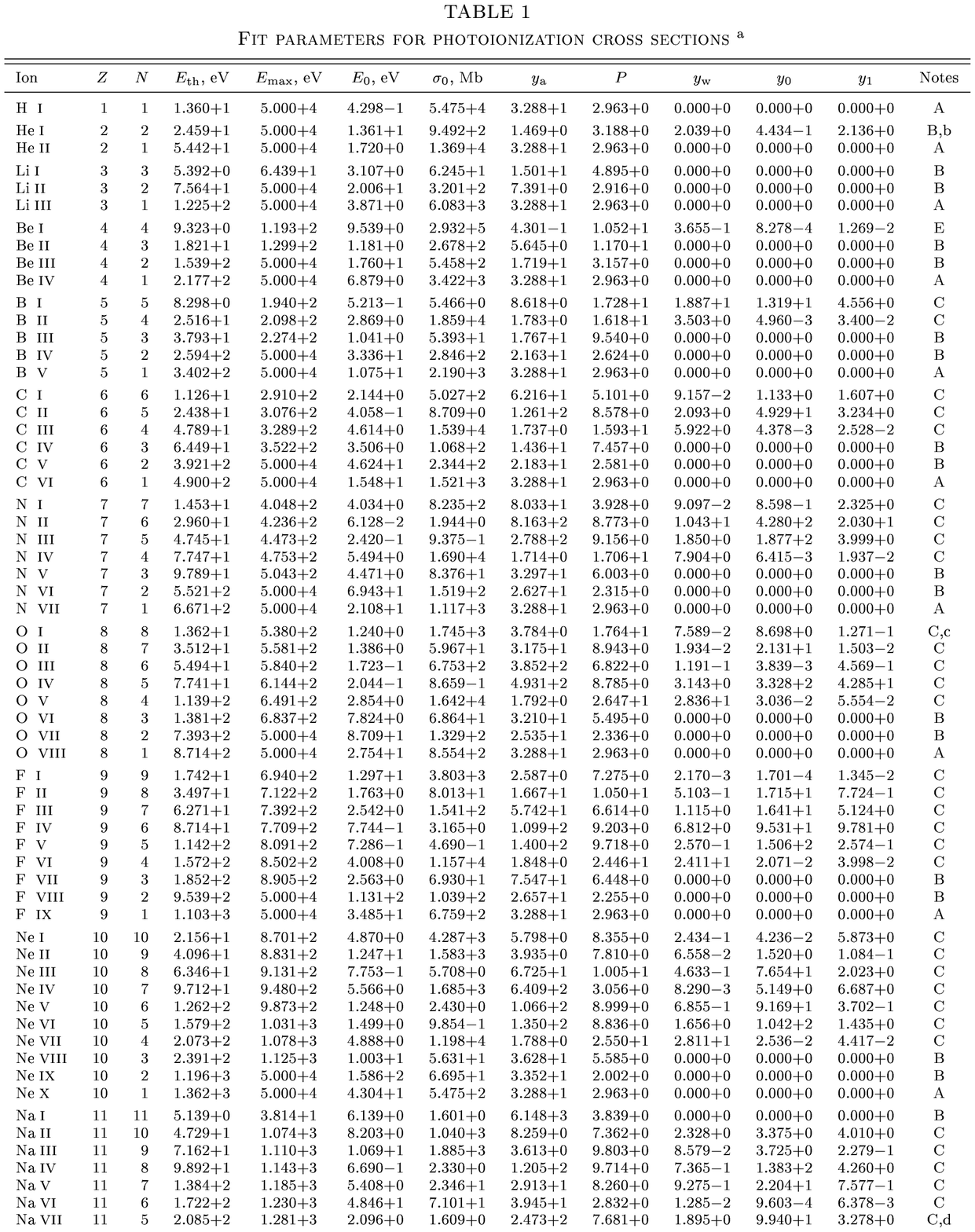}
\end{figure}

\begin{figure}
\plotone{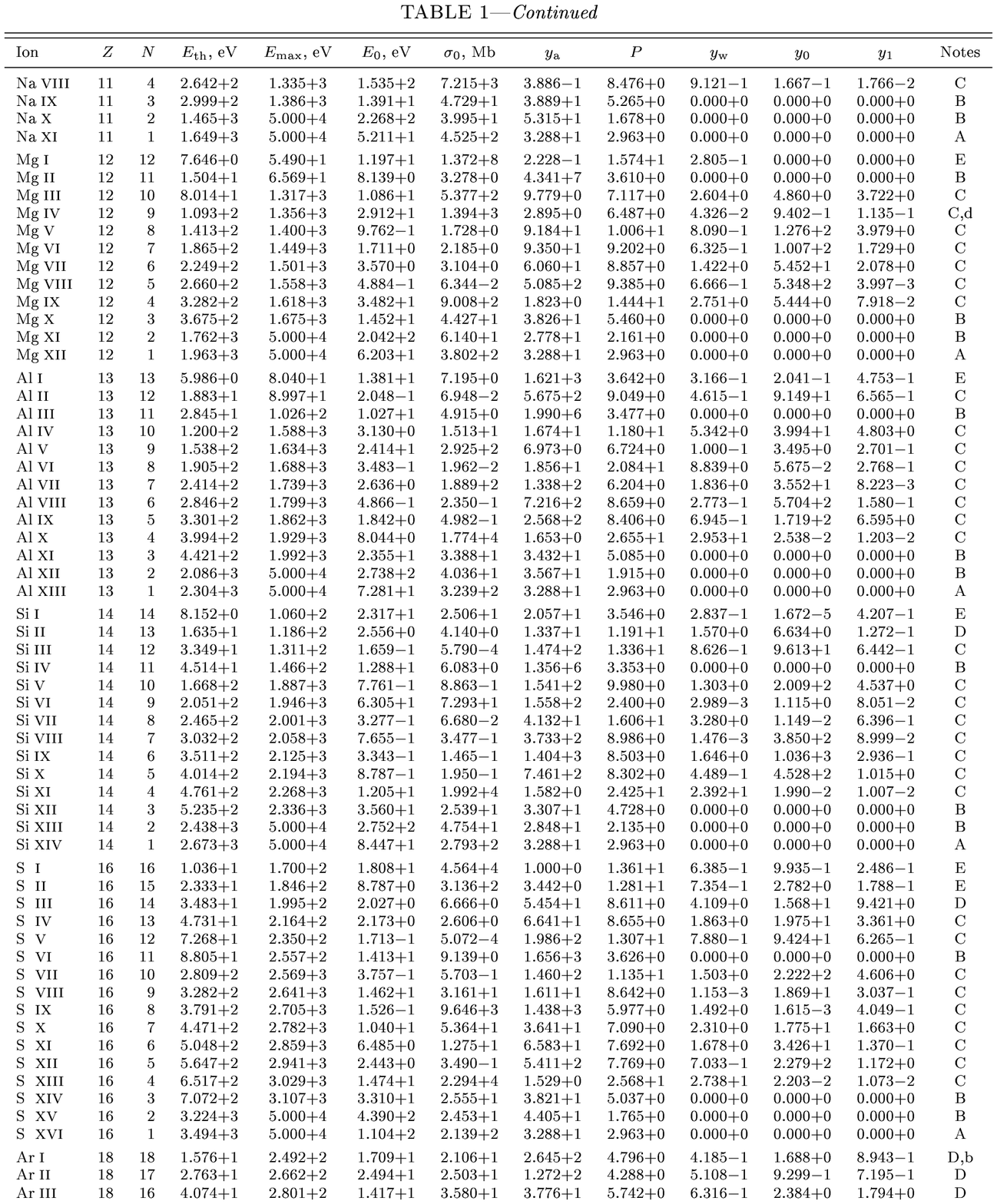}
\end{figure}

\begin{figure}
\plotone{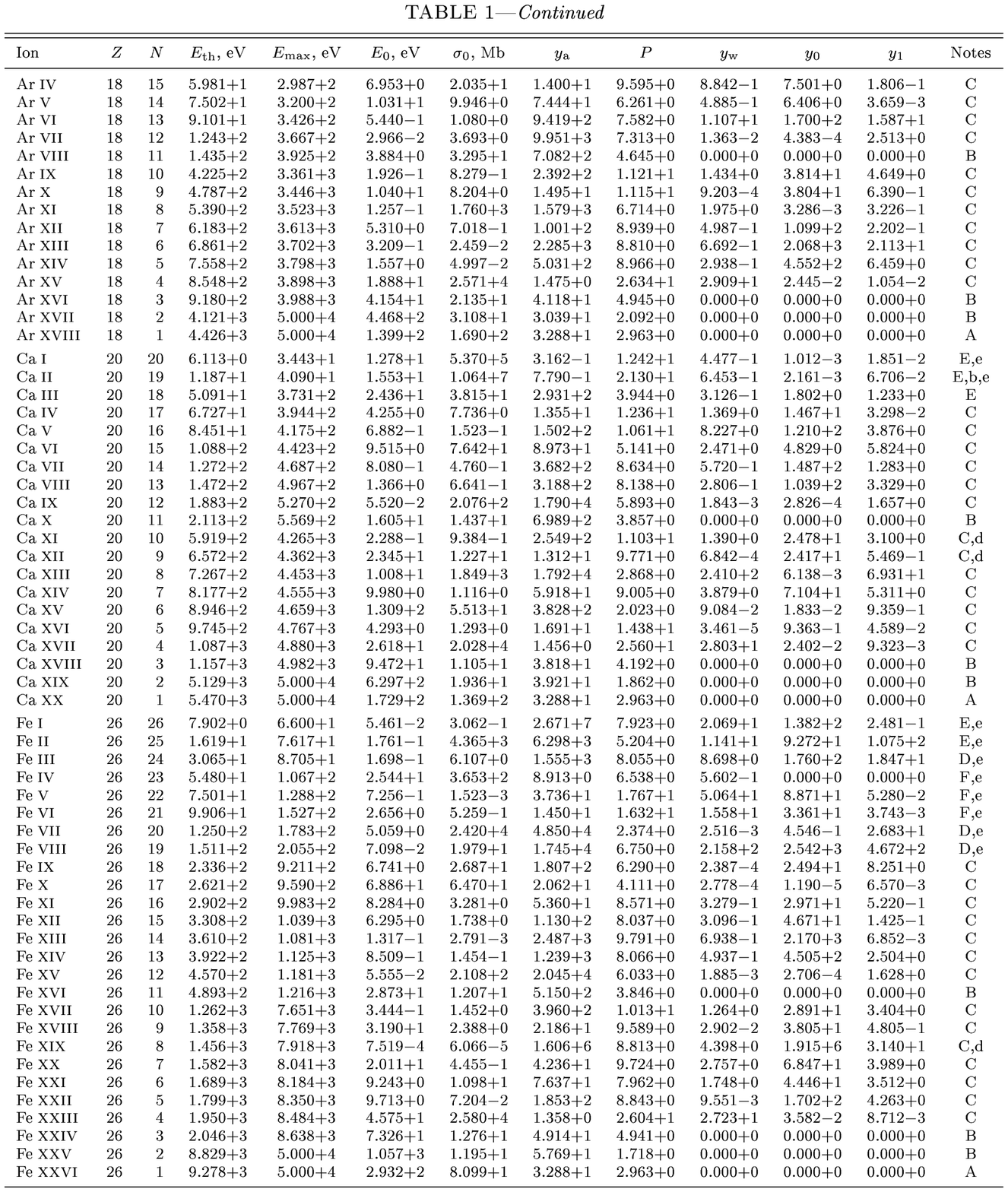}
\end{figure}

\begin{figure}
\plotone{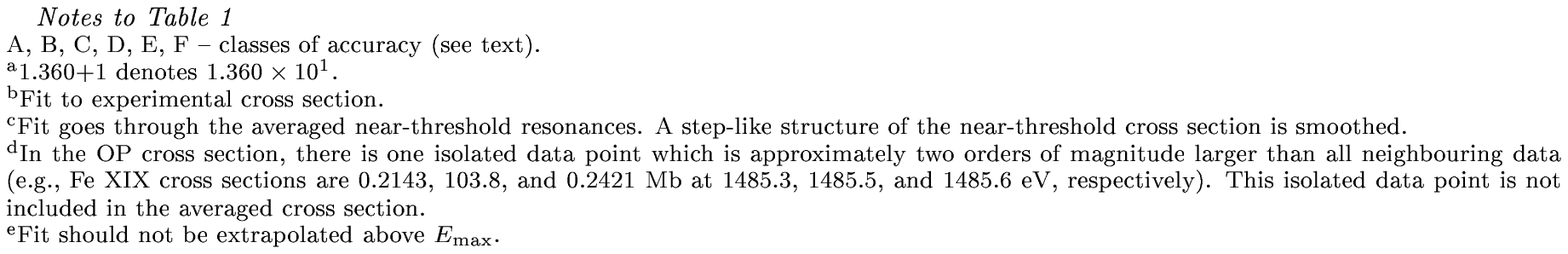}
\end{figure}


\clearpage
\begin{references}
\reference{baut1995} 
Bautista, M. A., \& Pradhan, A. K. 1995, J. Phys. B, 28, L173
\reference{butl1993} 
Butler, K., Mendoza, C., \& Zeippen, C. J. 1993, J. Phys. B, 26, 4409
\reference{cunt1993} 
Cunto, W., Mendoza, C., Ochsenbein, F., \& Zeippen, C.J., 1993, \aap, 275, L5
\reference{fern1987} 
Fernley, J. A., Taylor, K. T., \& Seaton, M. J. 1987, J. Phys. B, 20, 6457
\reference{hibb1994} Hibbert, A., \& Scott, M. P. 1994, J. Phys. B, 27, 1315
\reference{kaas1993} Kaastra, J. S., \& Mewe, R. 1993, \aaps, 97, 443
\reference{kell1987} Kelly, R. L. 1987, J. Phys. Chem. Ref. Data, 16, Suppl. 1
\reference{luo1989} Luo, D., \& Pradhan, A. K. 1989, J. Phys. B, 22, 3377
\reference{lyon1987} 
Lyon, I. C., Peart, B., Dolder, K., \& West, J. B. 1987, J. Phys. B, 20, 1471
\reference{marr1976} 
Marr, G. V., \& West, J. B. 1976, Atomic Data Nucl. Data Tables, 18, 497
\reference{mend1995} Mendoza, C., Eissner, W., Le Dourneuf, M., \& Zeippen, 
C. J. 1995, J. Phys. B, 28, 3485 
\reference{naha1993} Nahar, S. N., \& Pradhan, A. K. 1993, J. Phys. B, 26, 1109
\reference{naha1994} Nahar, S. N., \& Pradhan, A. K. 1994, J. Phys. B, 27, 429
\reference{peac1988} 
Peach, G., Saraph, H. E., \& Seaton, M. J. 1988, J. Phys. B, 21, 3669
\reference{reil1979} Reilman, R. F., \& Manson, S. T. 1979, \apjs, 40, 815
\reference{sams1990} 
Samson, J. A. R., \& Angel, G. C. 1990, Phys. Rev. A, 42, 1307
\reference{sams1991} Samson, J. A. R., Lyn, L., Haddad, G. N., \& Angel, 
G. C. 1991, J. de Physique IV Colloq., 1, C1-99
\reference{sams1994} Samson, J. A. R., He, Z. X., Yin, L., \& Haddad, G. N. 
1994, J. Phys. B, 27, 887
\reference{sara1992} 
Saraph, H. E., Storey, P. J., \& Taylor, K. T. 1992, J. Phys. B, 25, 4409
\reference{sawe1992} 
Sawey, P. M. J., \& Berrington, K. A. 1992, J. Phys. B, 25, 1451
\reference{seat1992} 
Seaton, M. J., et al. 1992, Rev. Mex. Astron. Astrofis., 23, 19
\reference{tull1990} 
Tully, J. A., Seaton, M. J., \& Berrington, K. A. 1990, J. Phys. B, 23, 3811
\reference{vern1993} Verner, D. A., Yakovlev, D. G., Band, I. M.,
\& Trzhaskovskaya, M. B. 1993, Atomic Data Nucl. Data Tables, 55, 233 (Paper I)
\reference{vern1994} 
Verner, D. A., Barthel, P. D., \& Tytler, D. 1994, \aaps, 108, 287
\reference{vern1995} 
Verner, D. A., \& Yakovlev, D. G. 1995, \aaps, 109, 125 (Paper II)
\end{references}
\end{document}